\documentclass[aps,epsfig,float,prl,twocolumn]{revtex4}
\usepackage{graphicx,amsmath}
\begin{document}
\title{Efficiency of thin film photocells}
\author{D. Mozyrsky and I. Martin}
\address{Theoretical Division, Los Alamos National Laboratory, Los Alamos, NM 87545, USA}
\date{Printed \today}
\begin{abstract}
We propose a new concept for the design of high-efficiency photocells based on
ultra-thin (submicron) semiconductor films of controlled thickness. Using a
microscopic model of a thin dielectric layer interacting with incident
electromagnetic radiation we evaluate the efficiency of conversion of solar
radiation into the electric power.  We determine the optimal range of
parameters which maximize the efficiency of such photovoltaic element.
\end{abstract}

\maketitle

Improvement of efficiency of semiconductor photovoltaic elements (solar cells)
has been an important technological challenge for several decades.  The maximum
possible efficiency obtains when every incident photon generates an
electron-hole pair, which then separates into electron flowing to cathode and
hole flowing to anode\cite{sq}. The limitations that reduce the efficiency of
the practical solar cells relative to the ideal are 1) light reflection at the
interfaces, 2) incomplete absorption of light entering the device due to finite
thickness, 3) electron-hole relaxation inside the absorbing medium during
diffusion to the leads\cite{b1,b2}. Interplay between two latter mechanisms leads to an
existence of an optimal device thickness, typically a few optical wavelengths.
Here we show that the interface reflection, commonly considered a completely
independent loss mechanism, shows an interesting interplay with absorption in
ultra-thin film devices.  This opens a possibility for a new generation of
ultra-thin (sub-wavelength) photovoltaic elements with efficiencies rivaling
the best conventional devices.

A ``working body'' of a solar cell is typically a semiconductor with relatively
high absorption index at frequencies corresponding to those of the sun quanta
$\hbar\omega_{\rm sun}\sim k_B T_{\rm sun}$, $T_{\rm sun}\simeq 6000\ K$.  Such
semiconductors, however, process a rather high refraction index $n$ at these
frequencies.  As a consequence, a fraction $(n-1)^2/(n+1)^2$ of the incident
light is reflected from the surface of the think device.  To reduce this loss,
often anti-reflective coating are applied to the surface of the device.  On the
other hand, for sub-wavelength thin films, the reflection can be significantly
smaller (for a reason similar to why even metallic films are transparent when
thin enough). Thus reducing the film thickness one should reach an optimum
where reflection is reduced but the absorption is still significant.

Also, in such thin devices the carrier recombination is naturally reduced.
Electron-hole recombination which prevents efficient charge separation in the
photocell is a major limiting factor in device operation. There are numerous
mechanisms which lead to the charge relaxation in a bulk of a semiconductor.
These mechanisms include spontaneous emission as well as phonon or impurity
induced relaxation. While it is difficult to control these processes in the
bulk of a semiconductor, it is clear that their contribution can be
significantly reduced if diffusion length of electrons and holes is large
compared to the width of the semiconducting layer.
\begin{figure}[h]
\vspace{-0 mm}
\centerline{\includegraphics[width=1.\columnwidth]{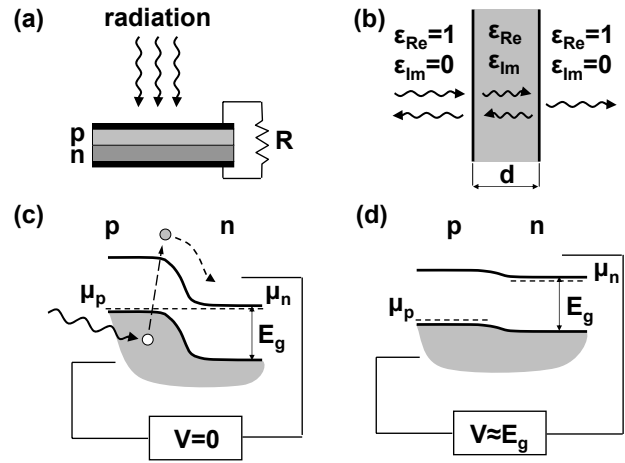}}
\caption{Insets (a) and (b): Schematics of the device. Insets (c)
and (d): Band structure of the device without and with external
load.} \vspace{-0 mm}
\end{figure}

Diffusion length for most semiconductors used in photocells is of the order of
a few microns. Since this distance is comparable to a typical wavelength of the
sunlight, one could expect that the specific absorption (the ratio of the
absorbed power to the incident power of radiation) in such a thin
semiconducting layer is insufficient for any practical use. In order to see
whether this is the case, it is instructive to look at the absorption of
radiation in a layer of thickness $d$, e.g. Fig. 1(a,b). For simplicity we
assume that the radiation is incident perpendicular to the surface of the layer
and is monochromatic with wavelength $\lambda$. The layer has a dielectric
constant whose real and imaginary parts are $\epsilon_{\rm Re}$ and
$\epsilon_{\rm Im}$ respectively. The specific absorption of the dielectric
layer can be easily evaluated by solving the wave equation $(n/c^2){\ddot A}
=\partial^2_z A$ ($A$ is radiation field vector-potential, $n=\epsilon^{1/2}$
is the refraction index, and $c$ is the speed of light) in the three regions,
i.g., Fig. 1(b), and taking into account the continuity conditions at the
boundaries of the dielectric layer, $A_1=A_2$, $A_2=A_3$, $\partial_z
A_1=\partial_z A_2$, etc. After straightforward algebra one finds that the
specific absorption is $P_{\rm abs} = 1-|t|^2-|r|^2$, where the amplitudes of
transmitted and reflected waves are
\begin{subequations}
\label{01}
\begin{eqnarray}
t ={4n\exp{({id/\lambda})}\over (n+1)^2\exp{({-ind/\lambda
})}-(n-1)^2\exp{({ind/\lambda})}}\,,\\
\label{01a} r={(n^2-1)[\exp{({ind/\lambda })}-\exp{({-ind/\lambda
})}]\over (n+1)^2\exp{({-ind/\lambda
})}-(n-1)^2\exp{({ind/\lambda})}}\, .\label{01b}
\end{eqnarray}
\end{subequations}
The specific absorption evaluated according to the above equations is presented
in Fig. 2 for a GaAs slab as a function of its thickness $d$. GaAs has a
relatively narrow ($\sim 1.4$ eV) bandgap and therefore is widely used in
high-efficiency photocells. Since dielectric function of GaAs is strongly
frequency dependent~\cite{r2}, in Fig. 2 we plot the specific absorption for
several energies typical to the quanta of solar radiation. The 2 eV curve
corresponds to the relatively low imaginary part of the dielectric constant and
thus saturates slowly exhibiting several oscillations due to the interference
between reflected and transmitted components. The 3 eV and 4 eV curves
correspond to much higher absorption (for example $\epsilon_{\rm Im}^{\rm
GaAs}(3\ eV) \simeq 17$) and saturate much faster. Prior to saturation both
curves exhibit a peak (again due to the interference) at roughly $d\simeq
\lambda/|\epsilon|$. Remarkably the value of the specific absorption at the
peak ($\sim 0.42$ at $d\simeq \lambda/|\epsilon|\simeq 20\ nm$ for 3 eV curve)
is nearly the same as its saturation value ($0.51$ at $d\gg \lambda$). Thus we
conclude that the solar radiation can be absorbed by a semiconducting layer of
{\it submicron} thickness almost as efficiently as by an infinitely thick slab.

In this paper, following the simple above considerations, we propose a new
concept for the design of photovoltaic elements based on thin semiconductor
films of controlled thickness. To put our arguments on more rigorous footings
in the following we consider a detailed microscopic model of a dielectric layer
interacting with solar radiation. Effective Hamiltonian can be derived from
standard quantum-mechanical interaction between matter and radiation,
$(e/mc){\bf A}({\bf r}){\bf\cdot p}+(e^2/2mc^2) {\bf A}^2(\bf r)$, where here
and in the following we assume Coulomb gauge for the electromagnetic field. The
radiation induces transitions between valence and conduction band of the
semiconductor. We assume that the temperature of the semiconductor is $0$, and
therefore these are the only possible transitions in the system (the valence
band is completely full and the conduction band is empty). Denoting the Bloch
states for the valence(conduction) bands as $\psi_{\bf k}^{v(c)}({\bf
r})=\exp{(-i{\bf k r})}u_{\bf k}^{v(c)}({\bf r})$, $H_0\psi_{\bf k}^{v(c)} =
E_{\bf k}^{v(c)}\psi_{\bf k}^{v(c)}$ ($H_0$ is a Hamiltonian of the crystal in
the absence of coupling to radiation field) one can rewrite the
radiation-matter interaction Hamiltonian in terms of single particle states in
the semiconductor as
\begin{equation}
H_{\rm int} = {e\over mc}\sum_{{\bf k}\,{\bf q}_\perp \alpha}
\langle u_{0}^c|p_\alpha|u_{0\alpha}^v\rangle \,c^\dagger_{\bf
k}\,d_{{\bf k-{\bf q}_\perp}\alpha}A_{{\bf
q}_\perp\alpha}^{z=0}+{\rm H.c.}\,. \label{1}
\end{equation}
We make the following assumptions: (1) While the electromagnetic field does not
significantly vary with distance inside the film, the electronic wave-functions
are effectively 3-dimensional - we assume that $\lambda_{\rm sun}\sim 2\pi\hbar
c/( k_BT_{\rm sun})\gg d\gg \hbar/(m^\ast E_g)^{1/2}$, where $d$ is the
thickness of the film, $m^\ast$ is exciton effective mass, $(m^\ast)^{-1} =
m_v^{-1} +m_c^{-1}$, $m_{v(c)}$ are effective masses in valence and conduction
bands, and $E_g$ is the band-gap (in this paper we assume that bands have
extremuma at zero momentum). This assumption allows one to carry out an
analytic calculation with rather simple and transparent results. We will
discuss the validity of this approximation at the end of the paper; (2) The
bands have different symmetry, say $s$ and $p$, so index $\alpha$ denotes
angular momentum of an electron in $p$ band. Since wave-vectors of the incident
radiation are nearly perpendicular to the surface of the film and the film is
assumed infinite in $x-y$ dimension, only in-plane components of the angular
momentum ($\alpha = x, y$) are relevant in Eq.~(\ref{1}); (3) The coupling in
Eq.~(\ref{1}) is isotropic and the coupling constant $t_\alpha=\langle
u_{0}^c|e p_\alpha/(mc)|u_{0\alpha}^v\rangle\simeq E_g p/(cS^{1/2})$, where $p$
is the effective dipole moment per unit cell of the film and $S$ is the surface
area of the film. (4) Due to external electric load the bands have effectively
different chemical potentials, $\mu_n$ and $\mu_p$, e.g., Figs.~ 1(c) and 1(d).
That is, once electron is promoted from valence to conduction band, it
immediately ``rolls over'' to the left lead, which corresponds to the infinite
transition rate between the semiconductor and the metallic lead (an infinitely
thin Shotkey barrier). Clearly, were the rate comparable or slower than the
electron-hole relaxation rate, the efficiency of the cell would have decreased.
(5) The $p-n$ junction is prepared (doped) as shown in Fig.~1(c) - the top of
the valence band in the $p$-doped area of the junction lies just below the
bottom of the of the conduction band in the $n$-doped area. Therefore the
maximum voltage the photo-cell can sustain is equal to $E_g/e$, which
corresponds to the assumption of maximum efficiency of Shockley and
Quasser\cite{sq}, i.e., each electronic transition from valence to conduction
band generates energy $E_g$ in the circuit.
\begin{figure}[h]
\vspace{-0 mm}
\centerline{\includegraphics[width=1.\columnwidth]{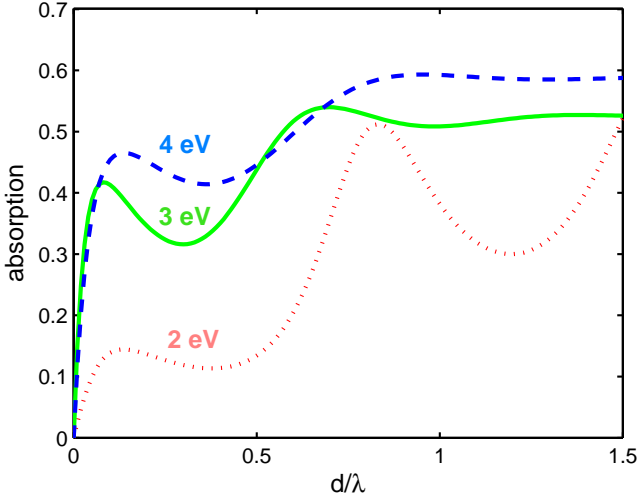}}
\caption{Specific absorption of GaAs slab as a function of its
thickness. Different curves correspond to different frequencies
(energies) of incident radiation.} \vspace{-0 mm}
\end{figure}
The photocurrent is defined as the rate of the charge transfer
between the valence and conduction band, ${\hat I}_{\rm ph} =
[H_{\rm int},\sum_{\bf k} c_{\bf k}^\dag c_{\bf k}]$. To the
lowest non-vanishing order it can be expressed as
\begin{equation}
I_{\rm ph} = {ie\over \hbar}\sum_{{\bf k}\,{\bf q}_\perp \alpha} t^2_\alpha
D^<_{\alpha\alpha{\bf q}_\perp}(E_{\bf k}^c - E_{{\bf k}-{\bf q}_\perp
\alpha}^v, z=z^\prime=0)\,. \label{2}
\end{equation}
The photocurrent of Eq.~(\ref{2}) is independent of the voltage across the cell
as far as it does not exceed the bandgap of the semiconductor, e.g., Fig. 1(d).
For larger bias, within our assumptions, a reverse current begins to flow. In
Eq.~(\ref{2}) $D^<_{\alpha\beta{\bf q}_\perp}(\omega, z,z^\prime) = \int dt
d^2{\bf r}_\perp \exp{i(\omega t+{\bf r}_\perp\cdot {\bf q}_\perp)} \,\langle
{\cal T}_K A^-_\alpha(t,{\bf r}_\perp,z) A^+_\beta(0,{\bf 0 },z^\prime)\rangle$
is the ``lesser'' Green's function of the electromagnetic field defined along
the Keldysh contour, where superscripts $\pm$ denote forward and return
branches of the contour~\cite{r3}. The Green's function is inhomogeneous along
$z$-direction, i.e., perpendicular to the film surface.

In order to incorporate effects of absorption and reflection from
the film, it is necessary to include renormalization of the photon
Green's function due to interactions with the film according to
Eq.~(\ref{1}). These effects can be treated by means of Dyson
equation, which, for the case of two-dimensional film reads
\begin{eqnarray}
&&{\hat D}_{\alpha\beta}(\omega, {\bf q}_\perp, z, z^\prime) =
{\hat D}_{0\,\alpha\beta}(\omega, {\bf q}_\perp, z- z^\prime)
\nonumber
\\&& +{\hat D}_{0\,\alpha\gamma}(\omega, {\bf q}_\perp,z){\hat\Sigma}_{\gamma\delta}(\omega,{\bf
q}_\perp){\hat D}_{\delta\beta}(\omega, {\bf q}_\perp, 0,
z^\prime)\,, \label{3}
\end{eqnarray}
where hats denote the standard $2\times 2$ matrix structure of the
non-equilibrium Green's functions. The self-energy
${\hat\Sigma}_{\gamma\delta}$ in Eq.~(\ref{3}) is quasi-two-dimensional. Since
$l\gg \hbar/(m^\ast E_g)^{1/2}$, effects related to the finite width of the
semiconducting slab can be neglected and
${\hat\Sigma}_{\gamma\delta}(\omega,{\bf q}_\perp,0) =
\sum_{q_z}{\hat\Sigma}_{\gamma\delta}(\omega,{\bf q})$, where
${\hat\Sigma}_{\gamma\delta}(\omega,{\bf q})$ is the self-energy defined for
the bulk of the semiconductor. Moreover, due to $x-y$ symmetry
${\hat\Sigma}_{\gamma\delta}$ reduces to a $\delta_{\gamma\delta}{\hat\Sigma}$,
where ${\hat\Sigma}$ depends only on $|{\bf q}_\perp|$. Therefore we obtain a
closed form equation for ${\hat D}_{\alpha\beta{\bf q}_\perp}(\omega,
z=z^\prime=0)$:
\begin{equation}
{\hat D}_{\alpha\beta}(0,0) = {\hat D}_{0\,\alpha\beta}(0) + {\hat
D}_{0\,\alpha\gamma}(0){\hat\Sigma}{\hat
D}_{\gamma\beta}(0,0)\,,\label{34}
\end{equation}
where components of the bare Green's function ${\hat
D}_{0\,\alpha\beta}$ for solar radiation are:
\begin{subequations}
\label{35}
\begin{eqnarray}
&&D_{0\,\alpha\beta}^{R(A)}(\omega, {\bf q}_\perp,0)=\sum_{q_z}
{4\pi(\delta_{\alpha\beta}
-q_\alpha q_\beta/q^2)\over \omega^2 -\omega_{\bf q}^2 \pm i\delta}\,,\\
\label{35a} &&D_{0\,\alpha\beta}^{<}(\omega, {\bf
q}_\perp,0)=\sum_{q_z}{4\pi^2i\over \omega_{\bf
q}}(\delta_{\alpha\beta} -q_\alpha q_\beta/q^2)\nonumber
\\&&
\times\big[\delta(\omega -\omega_{\bf q}){\tilde n}_{\bf
q}-\delta(\omega +\omega_{\bf q})(1+{\tilde n}_{-\bf q})\big]\,
.\label{35b}
\end{eqnarray}
\end{subequations}
In Eq.~(\ref{35}) $\omega_{\bf q}=\hbar c|{\bf q}|$ and the
${\tilde n}_{\bf q}$ is the distribution function of solar
radiation. We assume that the incident radiation wavevectors are
uniformly distributed within a cone with an opening angle $2\phi$
($\phi\ll 1$). Moreover, in order for the incident power to be
maximum, we assume that the surface of the cell is perpendicular
to the cone's axis. Then ${\tilde n}_{\bf q} = n_{\bf
q}^B\theta(q_z)\theta(\phi q_z-|{\bf q}_\perp|)$, where $n_{\bf
q}^B$ is Bose distribution function with temperature $T_{\rm
sun}$.

Furthermore due to the homogeneity of the Green's function ${\hat
D}_{\alpha\beta{\bf q}_\perp}$ in $x-y$ plane one can seek for
solution of Eq.~(\ref{34}) in the form ${\hat
D}_{\alpha\beta}={\hat D}_1\delta_{\alpha\beta}+{\hat D}_2\,
q_\alpha q_\beta/q^2$, where ${\hat D}_{1(2)}$ are $2\times 2$
matrices in the Keldysh space, but depend only on the absolute
value of the wavevector ${\bf q}$. Substitution of this ansatz
into Eq.~(\ref{34}) yields two independent equations  for ${\hat
D}_1$ and for ${\hat D}_3={\hat D}_1 + {\hat D}_2$. After solving
those equations one finds
\begin{eqnarray}
&&D_{1(3)}^< = {
(D_{01(3)}^{R})^{-1}D_{01(3)}^<(D_{01(3)}^{A})^{-1}+\Sigma^<\over
[(D_{01(3)}^{R})^{-1}-\Sigma^R][(D_{01(3)}^{A})^{-1}-\Sigma^A]}\,,
\label{4}
\end{eqnarray}
where ${\hat D}_{01}$ is the diagonal part of ${\hat
D}_{0\alpha\beta}$, e.g., Eqs.~(\ref{35}), and ${\hat D}_{03}=
{\hat D}_{01} + {\hat D}_{02}$, where ${\hat D}_{02}$ is the
transverse part of ${\hat D}_{0\alpha\beta}$. $\Sigma^{R(A)}$ and
$\Sigma^<$ are retarded(advanced) and ``lesser'' parts of photon
self-energy. Also we find a standard expression for the
retarded(advanced) Green's functions of the radiation:
\begin{eqnarray}
&&D_{1(3)}^{R(A)}=\big[(D_{01(3)}^{R(A)})^{-1}-\Sigma^{R(A)}\big]^{-1}\,.
\label{41}
\end{eqnarray}

We can now evaluate the photo-current in Eq.~(\ref{2}) in terms of the Green's
function of Eq.~(\ref{4}). The first contribution is due to absorption of
incident radiation accompanied by transfer of electrons from valence to
conduction band. It comes from the first term in the numerator in the RHS of
Eq.~(\ref{4}). This term can also lead to the reverse current due to
spontaneous and stimulated emission, i.e., transitions from conduction band to
the valence band accompanied by creation of a real photon. It arises due to the
$\delta(\omega+\omega_{\bf q})$ term in Eq.~(\ref{35b}). This process is,
however, not allowed while the energy gap $E_g$ exceeds the applied voltage
$\mu_n-\mu_p$, e.g., Fig. 1(c,d). In this situation the cell becomes a
light-emitting diode, and, as was stated above, we are not interested in such
case in this paper. The second contribution comes from the $\Sigma^<$ term in
the RHS of Eq.~(\ref{4}). It corresponds to the emission of virtual quanta of
radiation by one electron-hole pair and their subsequent re-absorbtion by
another pair, resulting in an incoherent simultaneous transfer of two electrons
from conduction to valence band. This process gives a reverse contribution to
the current which, once again, is non-zero only in the ``light-emitting''
regime (or at high temperature).

The self-energies in Eqs.~(\ref{4},\ref{41}) can be evaluated in
terms of the electronic Green's functions. The leading
contribution comes from the conventional polarization diagram,
i.e., a convolution of two electronic Green's functions. For
non-equilibrium situation one obtains
\begin{eqnarray}
&&\Sigma^R(\omega,0) ={it^2\over 4\pi}\sum_{\bf k}\int
d\omega^\prime \Big[G_{v {\bf k}}^K(\omega+\omega^\prime)G_{c \bf
k}^A(\omega^\prime) \nonumber
\\&& ~~~~~~~+G_{v {\bf k}}^R(\omega+\omega^\prime)G_{c
{\bf k}}^K(\omega^\prime) + (v\leftrightarrow c)\Big]\, .\label{5}
\end{eqnarray}
and $\Sigma^A=(\Sigma^R)^\ast$. In Eq.~(\ref{5}) $G_{v(c)}^{R(A)}$ are retarded
(advanced) Green's functions of valence(conduction) electrons and $G_{v(c)}^K$
is the Keldysh Green's function. Note that in Eq.~(\ref{5}) we evaluated the
self-energy at zero wavevector, since photon wavevectors are small compared to
those of electrons, and therefore the self-energy ${\hat\Sigma}$ is weakly
dependent on ${\bf q}$ for direct bandgap materials.

The self-energies $\Sigma^{(R)A}$ can be easily evaluated for
non-interacting electrons. Then $G_{v(c){\bf k}}^{R(A)}(\omega) =
(\omega -E_{\bf k}^{v(c)}\pm i\delta)^{-1}$ and $G_{v(c){\bf
k}}^K(\omega)=(1-2n^F_{n(p){\bf k}})\delta(\omega -E_{\bf
k}^{v(c)})$, where $n^F_{n(p){\bf k}}$ are Fermi filling factors.
The valence and conduction electrons are assumed to have chemical
potentials corresponding to those of the two leads, $\mu_n$ and
$\mu_p$ respectively. For these Green's functions the imaginary
part of the self-energy yields:
\begin{eqnarray}
&&\Sigma^{R(A)}_{\rm Im}(\omega,0) = \pm {a E_g^{1/2}\over
2\pi}\Theta(\omega-E_g)\sqrt{\omega -E_g} \, ,\label{6}
\end{eqnarray}
where we have introduced a dimensionless parameter
$a=p^2dE_g^{3/2}(m^\ast)^{3/2}/(2^{1/2}\hbar^4 c)$. A similar calculation for
the real part of $\Sigma^{R(A)}$ yields an ultraviolet divergence. This,
however, is an artifact of our approximation of the infinitely thin absorbing
layer (similar problem occurs in the quantum electrodynamics treatment of
electron-photon interaction). In a proper microscopic theory this divergence in
the real part of the self-energy is exactly cancelled by the
$e^2A^2(r)/(2mc^2)$ term in the radiation-matter interaction Hamiltonian (the
frequency sum rule). Taking into account this cancellation is equivalent to
performing the Kramers-Kr\"onig transformation on the dielectric function
[${\rm Re} \
 \epsilon(\omega) = 4\pi c^2{\rm Re}\, \Sigma(\omega )/\omega^2$], rather than self-energy:

\begin{figure}[h]
\vspace{-0 mm}
\centerline{\includegraphics[width=1.\columnwidth]{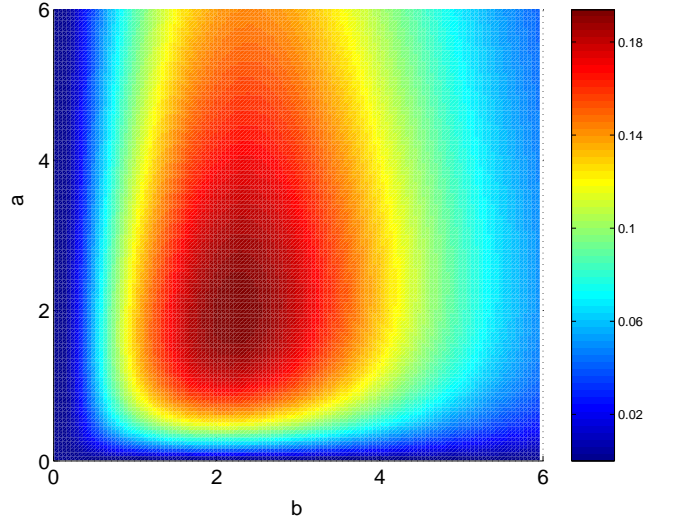}}
\caption{Dependence of photovoltaic efficiency on dimensionless
parameters $a$ and $b$.} \vspace{-0 mm}
\end{figure}

\begin{eqnarray}
&&\Sigma^{R}_{\rm Re}(\omega,0) = {\omega^2\over \pi}{\cal P}\int
{\Sigma^{R}_{\rm Im}(\omega,0) d\omega^\prime\over
{\omega^\prime}^2(\omega^\prime - \omega)}= {a E_g^{1/2}\over
2\pi}\nonumber\\
&&\times (2\sqrt{E_g}-\sqrt{|\omega + E_g|}-\sqrt{|\omega -
E_g|})\,,\label{8}
\end{eqnarray}
Note that in Eqs.~(\ref{6},\ref{8}) we used the three-dimensional expression
for the self-energy and therefore we can reexpress parameter $a$ in terms of
the conventional zero frequency dielectric constant $\epsilon_0$,
$a=2lE_g(\epsilon_0-1)/(\hbar c)$.

From Eqs.~(\ref{35},\ref{4},\ref{6},\ref{8}) one can evaluate the photocurrent
$I_{\rm ph}$ given by Eq.~(\ref{2}). In this paper we are interested in the
maximum efficiency of the photocell, which can be defined as $\eta_{\max} =
(IV)_{\max}/P_{\rm in}$, where $(IV)_{\max}$ is the power that is dissipated in
the circuit assuming an optimal load and $P_{\rm in}$ is the power of incident
solar radiation. Since the photo-current is only weakly dependent on voltage
for $V<E_g$ and becomes negative due to spontaneous and stimulated emission
emission for $V>E_g$, we have $(IV)_{\max}\simeq I_{\rm ph}E_g/e$. The incident
power per unit area $P_{\rm in}=cu$, where energy density $u=(2/v)\sum_{\bf q}
\hbar\omega_{\bf q}{\tilde n}_{\bf q}$, where $v$ is the mode quantization
volume and factor $2$ is due to two polarizations of the light waves. Carrying
out the calculation we obtain the following closed form expression for the
maximum photovoltaic efficiency of the cell:

\begin{eqnarray}
&&\eta_{\max} ={30ab^4\over \pi^4}\int_1^\infty {dx\over \exp{(bx)-1}}
\nonumber
\\&&\times{x^3\sqrt{x-1}\over (x+a\sqrt{x-1})^2+a^2(2-\sqrt{x+1}-\sqrt{x-1})^2}\,,~~~~\label{9}
\end{eqnarray}
where we introduced another dimensionless parameter $b=E_g/T_{\rm sun}$.
Function $\eta_{max}(a,b)$ shown in Figure 3 has a pronounced maximum at
$a\simeq 2.0$ and $b\simeq 2.4$. This maximum corresponds to the first maximum
of the specific absorption curves in Fig. 2. Note that since our microscopic
theory is valid only for thin dialectic layers ($d\ll \lambda_{\rm sun}$), it
does not account for the saturation of specific absorption at larger thickness,
i.e., when $d>\lambda$. However, since $\eta_{\max}$ reaches maximum at $d\sim
\lambda_{\rm sun}/\epsilon_0$, our theory is fully self-consistent for
semiconductors with sufficiently high value of the dielectric constant.

The optimal value of parameter $b$ corresponds to the bandgap energy $E_g^{\rm
opt}\simeq 1.2\ eV$. Since this value is close to the bandgap energy in GaAs
($E_g^{GaAs}\simeq 1.4 eV$), we conclude that GaAs is a good candidate for the
practical realization of thin film photocells. According to Eq.~(\ref{9}) the
optimal thickness of such GaAs layer is $d^{\rm opt}\simeq1.2\hbar c/[E_g^{\rm
opt}(\epsilon_0^{GaAs}-1)]\simeq 15\ nm$, which is within the fabrication
capabilities of contemporary molecular beam epitaxy technology.  Another
promising material is amorphous Silicone which, unlike crystalline Si, has
large imaginary part of dielectric constant\cite{a-Si}.  Moreover, in
ultra-thin devices considered here, the thermal equilibration of
photo-generated carriers may occur on the time scales longer than the charge
separation timescale.  Thus the hot-carrier physics\cite{hot} may lead to
further enhancement of the efficiency.

We thank D. Smith and A. Findikoglu for useful discussions. The work was
supported by the US DOE.

\end{document}